# Role of Post-Selection in Two-Photon Interferometer Tests of Bell's Inequality

J.D. Franson

*Physics Dept., University of Maryland, Baltimore County, Baltimore, MD, 21250*

Experimental tests of Bell's inequality based on two-photon interferometry consider only those coincidence events in which both photons arrive at the detectors at the same time. In a recent paper, Cabello et al. [Phys. Rev. Lett. **102**, 040401 (2009)] proposed a local model that is consistent with some aspects of these experiments. It is shown here that the model proposed by Cabello et al. is local but not realistic, and the role of post-selection in experiments of this kind is discussed.

In a recent paper [1], Cabello et al. presented a local model that is consistent with some aspects of the two-photon interferometer experiment [2] of Fig. 1. As a result, they concluded that experiments of this kind cannot rule out all local hidden-variable theories using the CHSH inequality. It will be shown here that the model presented by Cabello et al. is not a local realistic theory and it is noted that experiments of this kind have already been shown to be inconsistent with local realism using either chained Bell inequalities [3] or the original CHSH inequality [4].

Aerts et al. [3] have previously shown that "no local hidden variable model exists" for the two-photon interferometer of Fig. 1. Although their proof was based on the use of chained Bell inequalities and it requires measurements to be made at different phase settings from the usual CHSH inequality, they clearly showed that the output of the interferometer is inconsistent with local realism.

In the experiments of interest, two distant, identical interferometers have a short path S and a longer path L. Pairs of photons in an energy-time entangled state are incident on the two interferometers. Correlations are observed between those pairs of photons that arrive at the detectors at the same time (coincident detection), which means that both photons must have taken the longer paths (LL events) or both must have taken the shorter paths (SS events). Those correlations were shown [2] to violate Bell's inequality, but their interpretation is complicated by the fact that only the coincident events corresponding to LL or SS are accepted in a post-selection process.

The role of the post-selection can be understood by noting that the presence or absence of a single photon in either path through an interferometer can be predicted with certainty by placing a detector in the other, distant path, as has been demonstrated in numerous experiments. As a result, the path taken by a photon is an element of reality as defined by Einstein, Podolsky, and Rosen (EPR) [5] that must be specified by any local realistic theory while the photons are still traversing the interferometers. The LS and SL events do not contribute to the statistics collected and the remaining LL and SS events are described by a reduced set of hidden variables that must determine the results of the subsequent measurements without any further post-selection. As a result, it was previously shown [4] that the interferometer of Fig. 1 is inconsistent with any local hidden-variable theory, even when using the CHSH inequality.

Cabello et al. [1] proposed a local model that can duplicate some aspects of these experiments. In their model, the S/L decision "may be made as late as the detection time $t_D$" after the photons have already traversed the interferometer. But the presence or absence of a photon in either path can be predicted with certainty while the photons are still traversing the interferometer and this is an element of reality that must be specified in any local realistic theory. Cabello et al. explain this aspect of their model by noting that "the notion of EPR elements of reality was abandoned." As a result, their model is not a local realistic theory as defined by EPR [5]. Models of this kind have already been criticized for being inconsistent with other experiments [6].

It has been suggested that the proof of Ref. [4] should not be applied to the model of Cabello et al. on the grounds that Ref. [4] assumed rapidly-varying phase shifters, whereas the model of Cabello et al. was not intended to apply to that situation. As a result, it may be worthwhile to consider the need for rapidly-varying phase shifters in both the model of Ref. [1] as well as the proof in Ref. [4].

Within the context of local realistic theories, it seems reasonable to assume that the phase shifters are localized objects and that they can have no effect on the photons until the photons have propagated to the location of the phase shifters. This assumption was implicitly made by Cabello et al. [1] when they assumed that the S/L decision need not be made until the detection time $t_D$, which allows the photons to interact with the phase shifters before that decision is made. This makes their model more plausible, but a model in which the path taken by the photons is not determined until they have already traversed the interferometers is not a realistic theory as noted above. If the effects of the phase shifters are assumed to be localized in this way, then the proof of Ref. [4] does not require rapidly-varying phase settings and it applies directly to the model of Cabello et al.

Alternatively, one might argue that there could conceivably be some hypothetical and previously unobserved interaction that allows the settings of the phase shifters to somehow affect the path taken by the photons at the time they arrive at the beam splitters, despite the large distance between the two. Although a model of that kind does not seem very plausible, it would at least be a realistic model. Causality can be used to rule out possibilities of that kind if the phase settings are randomly chosen in less time than it would take for that information to reach the beam splitters at the speed of light [4], in analogy with the Aspect experiments [7]. It is important to note that randomly-chosen measurement settings with a space-like separation from the source of the entangled particles is required to rule out hypothetical interactions of this kind in any test of Bell's inequality, including the modified interferometer suggested by Cabello et al. In their case, the interferometer was extended back to the source, but the phase shifts would still have to be chosen in less time than it would take for information regarding the phase settings to propagate back to the beam splitters. As a result, Ref. [4] applies to the model of Cabello et al. regardless of whether or not we are willing to assume that the effects of the phase shifters are localized or that rapidly-varying settings of the phase shifters are required.

    In summary, energy-time entanglement violates local realism in the interferometer of Fig. 1 based either on chained Bell inequalities [3] or the original CHSH inequality [4]. Some aspects of these experiments are consistent with the model of Cabello et al., but only because that model abandons the notion of an EPR element of reality; their model is local but not realistic. Rapidly-varying settings of the measurement devices are required in any test of Bell's inequality if we desire to rule out implausible models in which the physical interactions associated with the measurement devices are not localized, but this situation is not unique to the interferometer of Fig. 1 or the proof of Ref. [4].

    This work was supported in part by the National Science Foundation (NSF) under grant 0652560.

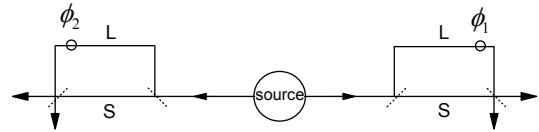

Fig. 1. A pair of energy-time entangled photons incident on two distant interferometers with a shorter path S and a longer path L [2]. Phase shifts $\phi_1$ and $\phi_2$ are inserted into the two longer paths. The output of this interferometer has been shown to be in disagreement with any local realistic theory [3, 4].

1. A. Cabello, A. Rossi, G. Vallone, F. De Martini, and P. Mataloni, Phys. Rev. Lett. **102**, 040401 (2009).
2. J.D. Franson, Phys. Rev. Lett. **62**, 2205-2208 (1989).
3. S. Aerts, P. Kwiat, J.-A. Larsson, and M. Zukowski, Phys. Rev. Lett. **83**, 2872 (1999).
4. J.D. Franson, Phys. Rev. A **61**, 012105 (1999).
5. A. Einstein, B. Podolsky, and N. Rosen, Phys. Rev. **41**, 777 (1935).
6. L.C. Ryff, Phys. Rev. Lett. **86**, 1908 (2001); preprint arXiv:0810.4825.
7. A. Aspect, J. Dalibard, and G. Roger, Phys. Rev. Lett. **49**, 1804 (1981).